\documentstyle[12pt,epsfig]{article}

\newlength{\dinwidth}
\newlength{\dinmargin}
\begin{document}
\def\bold#1{\setbox0=\hbox{$#1$}%
     \kern-.025em\copy0\kern-\wd0
     \kern.05em\copy0\kern-\wd0
     \kern-.025em\raise.0433em\box0 }
\def\slash#1{\setbox0=\hbox{$#1$}#1\hskip-\wd0\dimen0=5pt\advance
       \dimen0 by-\ht0\advance\dimen0 by\dp0\lower0.5\dimen0\hbox
         to\wd0{\hss\sl/\/\hss}}
\def\lq{\left [}
\def\rq{\right ]}
\def\LL{{\cal L}}
\def\VV{{\cal V}}
\def\AA{{\cal A}}
\def\BB{{\cal B}}
\def\MM{{\cal M}}
\def\ovl{\overline}
\newcommand{\be}{\begin{equation}}
\newcommand{\ee}{\end{equation}}
\newcommand{\bea}{\begin{eqnarray}}
\newcommand{\eea}{\end{eqnarray}}
\newcommand{\nn}{\nonumber}
\newcommand{\dd}{\displaystyle}
\newcommand{\bra}[1]{\left\langle #1 \right|}
\newcommand{\ket}[1]{\left| #1 \right\rangle}
\newcommand{\qq}{<0|{\bar q} q|0>}
\newcommand{\spur}[1]{\not\! #1 \,}
\thispagestyle{empty}
\vspace*{1cm}
\rightline{BARI-TH/97-266}
\rightline{May 1997}

\vspace*{3cm}

\begin{center}

{\LARGE \bf Violations of Local Duality \\ in the  Heavy Quark Sector\\}
\vspace*{1cm}
{\large P. Colangelo$^{a}$, C.A. Dominguez$^{b}$, G. Nardulli$^{a,c}$\\}
\vspace*{1.cm}

$^{a}$ Istituto Nazionale di Fisica Nucleare, Sezione di Bari, Italy\\
$^{b}$ Institute of Theoretical Physics and Astrophysics, University of 
Cape Town,\\Rondebosch 7700, South Africa\\
$^{c}$ Dipartimento di Fisica, Universit\'a di Bari, Italy\\
\end{center}

\vspace*{1cm}
\begin{abstract}
\noindent
We examine the origin of possible failures of the 
quark-hadron local duality. In particular,
we consider a correlator of two currents comprising heavy quark operators,
and compare the evaluation by the Operator Product Expansion with the
result obtained inserting
an infinite set of hadronic states, in the $N_c \to 
\infty$ limit. Whereas the smeared spectral functions agree with each other,
their local behaviour is different. The difference manifests itself in the 
form of a term ${\cal O} (\Lambda/\epsilon)$ ($\epsilon$ being the
residual energy)
that is not present in the 
expansion in powers of $\Lambda/\epsilon$ predicted by the OPE.
\end{abstract}

\newpage
\baselineskip=18pt
\setcounter{page}{1}
\section{Introduction}

Quark-gluon/hadron duality represents a key concept in the theoretical 
description of inclusive hadronic processes.
By its use, high energy 
processes such as, e.g.,  $e^+~e^-$ annihilation,
can be computed in terms of hadronic matrix elements of
operators in an expansion, the Operator Product Expansion (OPE), whose
leading term is represented by the perturbative Quantum-Chromo-Dynamics
expression. As a matter of fact, duality
provides the tool to extrapolate
from the deep Euclidean region, where the OPE is defined, to the 
Minkowski domain, where physical observables are measured. 
The basic assumption underlying this trade-off between 
hadronic quantities and quark-gluon amplitudes is that, at high energy
and/or momentum transfer, the hadronic behaviour should be well described
by quark and gluon interactions, provided that sufficiently 
inclusive variables are considered. 
\par
In the last few years 
this approach has been extended to the decays of hadrons containing one heavy
quark, exploiting the presence in these systems 
of a large parameter, the heavy quark mass
$m_Q$, that can be taken, at least  formally, infinitely large: $m_Q\to\infty$
\cite{c}. In this extension, however, a clear distinction must 
be maintained between semileptonic and nonleptonic decays. As 
a matter of fact, in general, the hadronic and the OPE amplitudes cannot 
be identical even at very high momentum transfer, due to the different structure
of their singularities: multiparticle thresholds in the former case and 
quark-gluon production thresholds in the latter. For semileptonic 
heavy hadron decays, however, this difficulty does not prevent duality 
to hold; as a matter of fact, in computing the semileptonic inclusive 
width, one has to integrate over lepton variables, which amounts 
to a smearing of the OPE width. The equality between smeared OPE
and physical, i.e. hadronic, widths is sometimes referred to as
{\it global duality} (assumed to be distinct 
from {\it local duality}, i.e. without smearing) and it is generally 
believed that global duality holds between quark-gluon and hadronic cross 
sections \cite{wein}; for $B$ and $\Lambda_b$ decays it 
has been proved explicitly in \cite{b} to two orders in the $1/m_Q$ expansion
and to the first order in $\alpha_s$, in the so-called small-velocity
(SV) limit ($m_b,m_c>>m_b-m_c>>\Lambda_{QCD})$ \cite{sv}.
\par
For nonleptonic heavy hadron decays, on the other hand, there are
no lepton momenta to be integrated and, therefore, one cannot use
global duality to prove the identification of OPE and hadronic observables. 
Therefore local duality must be assumed, which is a stronger hypothesis.
For this reason, the validity of OPE for the computation of
nonleptonic heavy hadron inclusive decays appears more debatable. 
For example in  \cite{a} it has been argued that the discrepancy between
the OPE prediction $\displaystyle{\frac{\tau(\Lambda_b)}{\tau(B_d)}>0.9}$
\footnote{For a discussion see \cite{neubert,colangelo,bigi}.} 
and the experimental result
 $\displaystyle{\frac{\tau(\Lambda_b)}{\tau(B_d)}=0.78\pm0.04}$ \cite{richman}
might be solved assuming a violation of local duality with the
appearance of a ${\cal O}(\frac{1}{m_Q})$ correction 
not predicted by OPE. In this context, it
may be useful to observe that in \cite{da} local duality has been proved for
nonleptonic $\Lambda_b$ and $B$ inclusive decays, for
the first two terms in the $1/m_b$ expansion, and at the order $\alpha_s$
in the perturbative expansion, assuming factorization of the weak
amplitudes. Also this result, however, holds in the SV limit, i.e. 
in a kinematical regime which is far off from the physical one; therefore, 
in absence of a general proof, the failure of local duality, with
the appearance of
${\cal O}(\frac{1}{m_Q})$ corrections to OPE, remains a logical possibility
and constitutes a simple explanation of the $\Lambda_b$ lifetime data.

It would be clearly extremely important to understand, in this scenario, 
the origin of this possible $\frac{1}{m_Q}$ correction to the
Operator Product Expansion, but this appears, at the moment, a formidable task
since,  at the
present stage of (analytical) understanding of non-perturbative
QCD, no first-principle answer can be given to the question of
how good duality really is and which violations it may suffer. A more modest,
but, nevertheless, potentially instructive aim could be to prove
the existence of a violation of local duality and the emergence of a
$\frac{1}{m_Q}$ correction in some definite model. This is 
the main purpose of the present letter.
\par
There have been some other recent studies on the validity of the 
quark-gluon/hadron duality in 
connection with non-perturbative QCD applications
in the heavy quark sector \cite{shif1,shif2}. 
Also these works are motivated by the hope to improve the precision with which 
various observables are determined using the
OPE coupled with duality. For example, the 
model studied in \cite{shif1} is based on an Ansatz for
the evaluation of the
two-point function involving the difference between scalar and pseudoscalar
heavy-light currents. In the chiral limit (for the light flavour) and the
infinite mass limit (for the heavy flavour), the coefficients of the OPE
for this correlator can be calculated analytically. A spectral function
model (in the time-like region) is then expanded in a power series 
(in the space-like region) and compared with the {\it exact} result. From
this comparison one can gauge the validity of (local) duality.
In this letter we consider the same two-point function and we analyze
two models: the first model, to be studied in the next Section, represents
a slight modification of \cite{shif1} and leads to a maximal violation 
of duality. The second model, discussed in Section 3,  uses
results for the hadronic evaluation of the two-point function
obtained by a constituent quark approach; therefore it
represents an improvement as compared to the previous Ans\"atze that are
{\it ad hoc} assumptions without physical justification. This calculation
explicitly shows a violation of local duality in the form
of an unexpected ${\cal O}(\frac{1}{m_Q})$ correction to OPE. 

\section{Mathematical models for the two-point function}

We begin by considering the following two-point function:
\be
\Pi (q)~=~\frac{i}{4} \int dx e^{i q x} <0|T(J_S(x)J_S^\dagger(0))-
T(J_P(x)J_P^\dagger(0))|0> = \Pi_S (q) - \Pi_P (q) \label{1}
\ee
where the scalar ($J_S$) and the pseudoscalar ($J_P$) currents are 
\be
J_S(x)~=~{\bar Q}(x) q(x) \label{2}
\ee
\be
J_P(x)~=~{\bar Q}(x) i \gamma_5 q(x) \label{3}
\ee
and $Q(x),~q(x)$ are heavy and light quark operators, respectively. This
two-point function is particularly simple, as in the chiral limit
$m_q \to 0$ it vanishes in perturbation theory. Hence,
$\Pi(q)$ is entirely non-perturbative. Additional simplifications take
place in the infinite mass limit $m_Q \to \infty$, where it is convenient
to write
\be
q^\mu~=~(m_Q - \epsilon, {\bf q} = 0) \; . \label{4}
\ee
In this limit the correlator becomes a function of $\epsilon$, i.e.
\be
\Pi(\epsilon)~=~\frac{1}{4} \int_0^{ + \infty} d \tau 
e^{- \epsilon \tau} \Phi (\tau)~~~~~~~~~(\epsilon>0) \label{5}
\ee
where $\Phi(\tau)$ is sometimes called the non-local quark condensate
\cite{rad}
\be
\Phi(\sqrt{-x^2})~=~<0|{\bar q}(x) e^{i g_s\int_0^x dy^\mu A_\mu(y)} q(0)|0>~. 
\label{6}\ee
In the limit $\epsilon >> \Lambda_{QCD}$, the OPE expression for
$\Pi(\epsilon)$ is given by
\be
\Pi_{OPE}(\epsilon) ~=~\frac{\qq}{4 \epsilon} \lq 1 - 
\frac{m_0^2}{8 \epsilon^2} + c_4 \frac{m_0^4}{\epsilon^4}
- c_6 \frac{m_0^6}{\epsilon^6}+...\rq ~, \label{7}
\ee
where $\qq=(-240 \; MeV)^3,~ 
{\dd{m_0^2=\frac{<0|g_s{\bar q} 
\sigma_{\mu\nu} G^{\mu\nu}  q|0>}{\qq}~=~0.8 \pm 0.2\; GeV^2}}$. 
The positive coefficients $c_{2 n}$  above, depend on the actual form of 
the non local condensate. We note explicitly the alternating signs in 
Eq.(\ref{7})  and the absence of even powers of $\epsilon^{-1}$; both these
features are consequences of the general principles on which the OPE is based 
\cite{shif1}.
In particular, the absence of a $D=4$ term, proportional to the gluon 
condensate $<\alpha_s G^2>$, is due to the limit $m_Q\to \infty , m_q \to 0$.

Let us now briefly
review the model proposed in \cite{shif1}
 to analyze duality violations. It is
given by
\be
\Pi(\epsilon) = \frac{\qq}{4 \bar{\Lambda}}
\beta (\frac{\epsilon + \bar{\Lambda}}{2 \bar{\Lambda}})\;,\label{8}
\ee
where $\bar \Lambda$ is a parameter and
\be
\beta(z) = \frac{1}{2} \; \left[ \psi \left( \frac{z+1}{2} \right) -
\psi \left( \frac{z}{2} \right) \right] \;\;, \label{9}
\ee
$\psi(z)$ being the logarithmic derivative of the Gamma function. The
model correlator admits the following series expansion
\be
\Pi (\epsilon) = \frac{\qq}{2 \bar{\Lambda}}\sum_{j=0}^{\infty} \;
\frac{(-1)^{j}}{\lambda \epsilon + 2j + 1} \label{10}
\ee
where $\lambda = 1/\bar{\Lambda}$. From this expression for $\Pi(\epsilon)$
we get, by the Mellin transform,
\be
\Phi(\tau) = \frac{\qq}{cosh(\bar{\Lambda} \tau)} \;.\label{11}
\ee
For $\epsilon \to \infty$, $\Pi(\epsilon)$ has the asymptotic expansion
\be
 \Pi (\epsilon)  \;
\sim_{\hskip -0.5cm \vspace*{0.75cm}_{\epsilon\to\infty}} \; 
\frac{\qq}{4 \epsilon} \;
\sum_{n=0}^{\infty} \; E_{2n}\frac{\bar{\Lambda}^{2n}}{\epsilon^{2n}} 
\label{12}
\ee
where $E_{2n}$ are Euler numbers ($E_0=1, E_2=-1, E_4=5, 
E_6=-61,...$)\cite{ab}.

Comparison of this result with
that of the OPE, Eq.(\ref{7}), indicates that this model is able to reproduce
the right power structure of $\Pi(\epsilon)$. The associated spectral
density corresponds to an infinite number of equally spaced poles located
along the negative $\epsilon$ axis, these poles having residues alternating
in sign. After smearing, the spectral function looks rather reasonable
\cite{shif1}.
However it is easy to show that this result strongly depends on the
Ansatz (\ref{8}). As a matter of fact, let us consider 
the slightly modified model
\be
\Pi (\epsilon) = \mbox{const.} \times  \sum_{j=1}^{\infty} \;
\frac{(-1)^{j}}{\lambda \epsilon + 2j } \label{13}
\ee
instead of (\ref{10}). This implies
\be
\Phi(\tau) = \mbox{const.} \times [tanh (\frac{\tau}{\lambda}) - 1] \;.
\label{14}
\ee
In the limit $\epsilon \to \infty$, this $\Pi(\epsilon)$ becomes
\be
 \Pi (\epsilon)  \;
\sim_{\hskip -0.5cm \vspace*{0.75cm}_{\epsilon\to\infty}} \; 
\mbox{const.} \times \frac{1}{\epsilon} \; \left[ 1 +
\sum_{n=0}^{\infty} \; \frac{C_{2n+1}}{\epsilon^{2n+1}} \right]~.
\label{15}
\ee
A comparison with the OPE result, Eq.(\ref{7}), shows that except for the first
term, the power structure is wrong, so that one obtains a maximal violation
of duality. This model is still quite realistic. The spectral function
is also made of an infinite number of zero-width resonances, with
alternating sign residues, and located along the negative $\epsilon$
axis. The only difference with the model of \cite{shif1} is a shift in the 
location of the poles, due to the absence of the unit factor in the
denominator of Eq.(\ref{13}).
\par
From this example we learn that the agreement found in 
\cite{shif1}  between
the hadronic evaluation of the correlator and the OPE result
might be fortuitous; in order to get a deeper understanding we compute 
now the correlator Eq.(\ref{1}) using a more realistic
model for the spectral density.

\section{A realistic model for the hadronic correlator}

We evaluate the two-point function (\ref{1})
by inserting hadronic states between the currents. In general this can only be 
done in some approximation: we choose to insert an infinite number 
of states, but in the $N_c \to \infty$ limit ($N_c=$ number of colours),
where the surviving contributions are $J^P=0^+$ and $0^-$
single particle states,
contributing respectively
to the scalar ($\Pi_S$) and pseudoscalar ($\Pi_P$) part of $\Pi$. By
denoting $|S_n>$ and $|P_n>$  these states, with masses
$M_{S_n}$ and $M_{P_n}$  respectively, we define the current-particle
matrix-elements:
\bea
<0|J_S|S_n>&=&\frac{M^2_{S_n}}{m_Q} f_{S_n}  \\
<0|J_P|P_n>&=&\frac{M^2_{P_n}}{m_Q} f_{P_n}~. 
\eea
In the $m_Q \to \infty$ limit we have
\bea
M_{S_n}&=&m_Q+\delta_{S_n} + {\cal O}(\frac{1}{m_Q}) \label{18} \\  
M_{P_n}&=&m_Q+\delta_{P_n} + {\cal O}(\frac{1}{m_Q})~.\label{19}  
\eea

The binding energies  $\delta_{S_n} $ and $\delta_{P_n}$ can be obtained by 
solving the wave equation
\be
\lq \sqrt{- \nabla^2+m^2_Q}+
\sqrt{- \nabla^2+m^2_q} + V({\vec r})\rq \Psi_n({\vec r})=M_n \Psi_n({\vec r})
\ee
in the limit $m_Q\to \infty, m_q \to 0$. Assuming a central potential
$V({\vec r})=V(r)$ we can write $\Psi_n({\vec r})=
Y_{{\ell}{m}}({\hat r}) 
\frac{u_{\ell}^{(n)}(r)}{r}$. The scalar particles are obtained for
${\ell}=1$ ($P$-waves) and the pseudoscalar particles for ${\ell}=0$
($S-$waves). The corresponding equation \cite{piet} for 
$\delta_{S_n}$, $\delta_{P_n}$ is given, in this limit, by
\be
\lq V(r) - \delta_{\ell}^{(n)} \rq u_{\ell}^{(n)}(r) +\frac{2}{\pi}
\int_0^\infty dr^\prime \int_0^\infty dk k \chi_{\ell}(kr)  
\chi_{\ell}(kr^\prime)  u_{\ell}^{(n)}(r^\prime)~=0~ 
\ee
where
the relation between $\delta_{S_n},\delta_{P_n}$ in
(\ref{18},\ref{19}) and   $\delta_{\ell}^{(n)}$ is as follows:
 $\delta_0^{(n)}=\delta_{P_n},~\delta_1^{(n)}=\delta_{S_n}$; moreover,
$\chi_{\ell}(x)=x j_{\ell}(x) ~~(j_{\ell}$ are the spherical Bessel
functions). 

$V(r)$ is the static interquark potential; several forms have been 
studied in the literature which reproduce the experimental spectrum 
of the heavy $Q{\bar q}$ mesons. In general they behave linearly $V(r) 
\simeq \mu^2 r$ for $r \to \infty$ 
and have a coulombic behaviour $(V(r)\simeq 
\frac{\alpha_s}{r})$ at small distances. In order to
simplify  our discussion we assume
\be
V(r)=\mu^2 r 
\ee
with constant $\mu$ (string tension). This form is adequate for our
aims because the coulombic part of 
the potential mainly affects the first states ($n=0$) 
and becomes negligible with increasing $n$. 
\footnote{For example,
using the Richardson potential and solving numerically the wave equation
\cite{piet} one obtains a
coulombic correction to the meson mass as follows: 
$\Delta M_n/M_n=7.9 \%, 5.6 \%$ and $4.4 \%$ 
for the radial quantum number $n=0,1,2$ respectively. 
}.
\par
In order to solve Eq.(21) for any integer $n$ ($n=0,1...$), 
we work in the WKB approximation \cite{prep}. The  spectrum 
and the wavefunctions are obtained by the usual WKB procedure 
\cite{piet,prep}; the spectrum is given by
\be
\delta_{\ell}^{(n)}~=~\mu {\sqrt{\pi(2 n + \ell + \frac{3}{2}) }} \;\; ;
\ee
the wave function, for $r \leq r_0 = \frac{\delta_{\ell}^{(n)}}{\mu^2}$,
reads as follows:
\be 
u_{\ell}^{(n)}(r)~=~\eta \sqrt{m_Q}\;\;\chi_{\ell}\lq \delta_{\ell}^{(n)} r
-\frac{\mu^2 r^2}{2}\rq~~~~~~~~~~(r \leq r_0) \;\;\;,
\ee
whereas for larger values of $r$ it decreases  exponentially. Let us observe
explicitly that the factor $\sqrt{m_Q}$
arises from the covariant normalization condition
\be
\int d {\vec r} \;\; |\Psi_n({\vec r} )|^2 = 2 M_n \;\;\;. 
\ee
\par
Let us now turn to the coupling $f_{S_n}=f_1^{(n)}$ and  $f_{P_n}=f_0^{(n)}$.
From the expression\cite{cea}
\be
f_{\ell}^{(n)}= \sqrt{\frac{3}{2}} \frac{1}{\pi M_{n,{\ell} }}
\int_0^\infty d k ~ k~ {\tilde u}^{(n)}_{\ell}(k) \;\;\;, 
\ee
with the function ${\tilde u}^{(n)}_{\ell}(k)$ defined as
\be
{\tilde u}^{(n)}_{\ell}(k) = \int_0^\infty dr~ \chi_{\ell}(kr) 
u_{\ell}(r) ~~,
\ee
one obtains
\be
f_{\ell}^{(n)}=\sqrt{ \frac{ 3 m_Q \delta_{\ell}^{(n)} } {\pi} } 
\frac{\mu}{M_{n,{\ell}}}~~. \label{28}
\ee
Let us now reconsider the correlator Eq.(\ref{1}).
In the $N_c\to \infty$ limit the products of currents in Eq.(1)
are saturated by single particle states with $J^P=0^+$ and $0^-$, respectively.
Therefore one can write:
\be
\Pi^{had}(\epsilon) = \frac{1}{8 m_Q^3}\sum_{n=0}^\infty 
\lq \frac{ \left [f_1^{(n)}\rq^2}{\epsilon +\delta_1^{(n)}} - 
\frac{\left[ f_0^{(n)}\right]^2}{\epsilon +\delta_0^{(n)}} \rq~~, 
\ee
which, in the limit $m_Q\to\infty$, becomes
\be
\Pi^{had}(\epsilon)=\frac{3 \mu^2}{8 \pi} \sum_{n=0}^\infty 
\lq \frac{\delta_1^{(n)}}{\epsilon +\delta_1^{(n)}} - 
\frac{\delta_0^{(n)}}{\epsilon +\delta_0^{(n)}} \rq
=\frac{3 \mu^3}{8 \sqrt{\pi}}\sum_{n=0}^\infty 
\frac{ (-1)^{n+1}\sqrt{n+3/2}}{\epsilon+\mu\sqrt{\pi(n+3/2)}}~~.\label{30}
\ee
\par
Before continuing in the analysis, a comment is in order on the model.
The choice of considering 
only single particle states between the currents in the correlator
(i.e. taking the $N_c\to\infty$ limit)
is not too restrictive, as it can be shown by considering, for example, 
the imaginary part of the correlator of scalar currents $\Pi_S$ in
Eq.(\ref{1}) (the same result is obtained taking $\Pi_P$). By computing 
the imaginary part of the quark loop diagram  one obtains,
for $E=- \epsilon>\to +\infty$ and at the leading order in $E$:
\be
Im \Pi_S^{OPE}(E) ~\to ~\frac{3 E^2}{8 \pi}~. \label{31} 
\ee
The resonance model gives
\bea
Im\Pi_S^{had}&=&Im\frac{1}{8}
\sum_n\frac{\lq f_1^{(n)}\rq^2 M_{n,1}}{-E+\delta_1^{(n)}-i\epsilon} \nn \\
&=& \frac{\pi}{8}\sum_n\lq f_1^{(n)}\rq^2 M_{n,1}\delta(E-\delta_1^{(n)}) 
\nn \\
&=& \frac{3 \mu^2 E}{8}\sum_{n=0}^\infty \delta \lq E-\mu\sqrt{2 \pi(n+7/4)}\rq
~~.
\eea
$Im~ \Pi_S^{had}(E)  $ looks very different  from $Im~ \Pi_S^{OPE}(E) $
in Eq.(\ref{31}), due in particular to the presence of the infinite set 
of Dirac $\delta$ functions in Eq.(32). On the other hand it is well known 
that a comparison should be made only after a smearing of these 
distributions \cite{wein}. A convenient way to 
do this in the present case is make the approximation
\be
\sum_n \to \int dn~~.
\ee
Hence, we obtain
\bea
Im~\Pi_S^{had}&=& \frac{3 \mu^2 E}{8}\int dn
\delta \lq E-\mu \sqrt{2 \pi(n+7/4)} \rq \nn \\
&=&\frac{3 E^2}{8 \pi}=Im\Pi_S^{OPE}(E)~. 
\eea
Therefore $\Pi_S(E)$  and  $\Pi_P(E)$  satisfy global duality at 
least at the leading order for $E\to\infty$.
\par
In the sequel we shall discuss the next-to-leading contributions that 
are responsible for the difference $\Pi=\Pi_S-\Pi_P$; for the time 
being let us comment on the value of the parameter $\mu$ and the 
approximate choice Eq.(22) for the potential.
Assuming that the WKB approximation works reasonably well already for
the the first quantum number $n=0$, we obtain
\be
\mu\simeq \sqrt{\frac{2}{3 \pi}}(m_B-m_b)~; \label{35}
\ee
for $m_b=4.6-4.7$ GeV this gives $\mu\simeq 300$ MeV.
The first ($n=0)$ meson state may be sensitive to the small 
distance, coulombic part of the interquark potential that we have 
omitted in Eq.(22); however, as far as we neglect consistently 
${\cal O}(\alpha_s)$ terms also in the OPE counterpart of $\Pi^{had}$,
this approximation is reasonable and the duality should be equally valid.
\par
Let us now apply the same resonance model
to the correlator 
$\Pi(q)=\Pi_S(q) - \Pi_P(q)$ in Eq.(\ref{1}). When we compute 
the correlator inserting hadronic
states, from Eq.(\ref{30}) we find $(E=- \epsilon>0)$:
\be
R^{had}(E)=Im \Pi^{had}(E)=\frac{3 \mu^2 E}{8}\sum_{n=0}^\infty (-1)^n
\delta \lq E- \mu \sqrt{\pi(n+\frac{3}{2})}\rq ~~. \label{36}
\ee
On the other hand, from the first four terms in Eq.(\ref{7}), we have:
\bea
R^{OPE} (E)&=&Im\Pi^{OPE} (E)\nn \\
= &-&\frac{\pi}{4}\qq\lq \delta(E)-\frac{m_0^2}{16}
\delta^{\prime \prime}(E)+ c_4\frac{m_0^4}{4!} \delta^{(IV)}(E)
-c_6\frac{m_0^6}{6!}\delta^{(VI)}(E)...\rq~; \label{37}
\eea
$c_4$ and $c_6$ are unknowns: in the model of \cite{shif1} they
are given by ${\dd c_4=\frac{5}{64},~c_6=\frac{61}{512}}.$
Also in the present case the two expressions look very 
different, but they  can be compared after an 
appropriate  smearing; following \cite{wein} we consider
\bea
{\bar R}^{had}(E,\Delta)&=&\frac{\Delta}{\pi}
\int dE^\prime~ \frac{ R^{had}(E^\prime)}{(E-E^\prime)^2+\Delta^2} \label{38}
 \\
{\bar R}^{OPE}(E,\Delta)&=&\frac{\Delta}{\pi}
\int dE^\prime~ \frac{ R^{OPE}(E^\prime)}{(E-E^\prime)^2+\Delta^2} \;\;\; .
\label{39}
\eea
In the limit $\Delta\to 0, {\bar R}(E,\Delta)\to R(E)$; however one has 
to impose $\Delta>>\Lambda_{QCD}$ \cite{wein}. 
As a matter of fact,
\be
{\bar R }(E,\Delta)=\frac{1}{2i}\lq\Pi(E-i \Delta)-\Pi(E+i\Delta)\rq
\label{40}
\ee
and, for $\Delta>>\Lambda_{QCD}$, we are far enough from the singularities
for OPE to be valid and duality to hold.
\par
For very large $E$ the two expressions should coincide regardless 
of $\Delta$. From previous equations one
finds that this request implies the relation
\be
\qq=-3 \frac{0.51}{2}\frac{\mu^3}{\sqrt{\pi}},\label{extra}
\ee
and, for $\qq=(-240 $MeV$)^3$, this implies $\mu=317$ MeV; it 
can be noted that this number agrees nicely with the previous result 
Eq.(\ref{35}). For smaller values of $E$, i.e. $1<E<20\; GeV$,
the numerical results concerning the comparison between
(\ref{38}) and (\ref{39}) are reported in Fig. 1, where we plot
the ratio:
\be
P=\frac{{\bar R}^{had}} {{\bar R}^{OPE}} \nonumber \nn \label{41}  
\ee
of the smeared quantities as a function of $E$ for
several values of $\Delta$. We observe that, as expected,
the agreement between ${\bar R}^{had}$ and ${\bar R}^{OPE}$ increases
with increasing $\Delta$; in
particular for $\Delta=3.0$ GeV the difference does not exceed $20\%$. 
In any case, for any value of the smearing parameter $\Delta$,  the 
ratio (\ref{41}) tends to unity for large $E$ values.
\par
A further test of the soundness of our approximation is provided
by the evaluation of
\be
{\hat F}= {\sqrt{m_b}} f_B
\ee
in the limit $m_b\to\infty$,
 using the WKB formula (\ref{28}) with $n=\ell =0$. 
We get ${\hat F}= 0.26$ GeV$^{3/2}$ to be compared with
the QCD sum rule result, obtained without $\alpha_s$ corrections, 
i.e. in the same approximation used in this work: ${\hat F}= 0.30\pm 0.05$ 
GeV$^{3/2}$\cite{norberto} (including  $\alpha_s$ corrections, 
one would  get  ${\hat F}= 0.41\pm 0.04$ GeV$^{3/2}$).

In conclusion we have tested that 
the resonance model we have proposed satisfies 
global duality: the smeared imaginary parts of the correlator, 
when computed by OPE or by hadron states agree with each other.
When we consider local duality, 
however, the expressions are significantly 
different. As a matter of fact, starting 
from Eq.(\ref{30}), we obtain, by the Mellin transform:
\be
{\tilde \Pi}(\tau)=\frac{1}{2\pi i}\int_{\sigma-i\infty}^{\sigma+i\infty}
d \epsilon~ e^{\epsilon \tau}\Pi(\epsilon)
\ee
with
\be
{\tilde \Pi}(\tau)=\frac{3\mu^3}{8\sqrt\pi}\sum_{n=0}^{\infty} (-1)^{n+1} 
\sqrt{\pi (n+3/2)} \; e^{-\tau\mu\sqrt{\pi (n+3/2)}}~.
\ee
${\tilde \Pi}$ can be expanded for small $\tau$; we find numerically
\be
{\tilde \Pi}(\tau)=\frac{3\mu^3}{8\sqrt\pi} \lq -0.51
+\frac{\mu \sqrt\pi\tau}{2}
-0.46\frac{\mu^2\pi \tau^2}{2}...\rq~~. 
\ee
Since
\be
\Pi(\epsilon)=\int_0^\infty d \tau e^{-\tau \epsilon} 
{\tilde \Pi}(\tau)~, 
\ee
we find
\be
\Pi^{had}(\epsilon)=-3~\frac{0.51 \mu^3}{8\sqrt{\pi}}\frac{1}{\epsilon}
\lq 1-\frac{{\tilde m}_0}{\epsilon}+0.93\frac{{\tilde m}_0^2}{\epsilon^2}
... \rq~.
\ee
The factor multiplying 
$\frac{1}{\epsilon}$ coincides numerically with $\frac{\qq}{4}$ 
for $\mu=317$ MeV (see eq.(\ref{extra})). As for 
the mass parameter ${\tilde m}_0$,
numerically we find ${\tilde m}_0=~560$ MeV. In conclusion we get 
($\epsilon<0$):
\be
\Pi^{had}(\epsilon)=\frac{\qq}{4\epsilon}
\lq 1-\frac{{\tilde m}_0}{\epsilon}+0.93\frac{{\tilde m}_0^2}{\epsilon^2}
+...\rq ~. \label{last}
\ee
A comparison between 
Eq.(\ref{7}) and Eq.(\ref{last}) shows a violation of local duality
which manifests itself in the form of an unexpected term in the 
${\dd \frac{1}{\epsilon}}$ asymptotic expansion.

\section{Conclusions}
Studying simple correlators of quark currents, we have
found, in a particular quark model and in the $N_c \to \infty$ limit,
an  explicit example of violation of local duality which occurs
in a condition where global duality is verified. The difference
between the hadronic and the OPE spectral functions is made evident by 
the  $\Lambda/\epsilon$ term which is absent in the expansion predicted by OPE.

The calculation in QCD of correlators of quark operators between
hadronic states, 
as required for the evaluation of, e.g., the $B_d$ and $\Lambda_b$ 
inclusive decay 
widths, is, of course, beyond the aims of the present analysis.
However, the example described in this paper supports the suggestion 
that the $\Lambda_b$ lifetime data could find
an explanation in the presence of a $1/m_Q$ correction not included in the 
usual QCD-OPE expansion. 

\vspace*{2cm}
{\noindent \bf Acknowledgments\\} 
\noindent 
We thank  N. Paver for useful discussions. The work of (CAD) has been supported
in part by the FRD (South Africa).

\newpage

\hskip 3 cm {\bf FIGURE CAPTIONS}
\vskip 1 cm
\noindent
{\bf Fig. 1} \par
\noindent
Plot of the ratio (\ref{41}) as a function of $E$ for
several values of $\Delta$ (from top to bottom:
$\Delta=1.5, 2.0, 3.0$ and $5.0$ GeV).
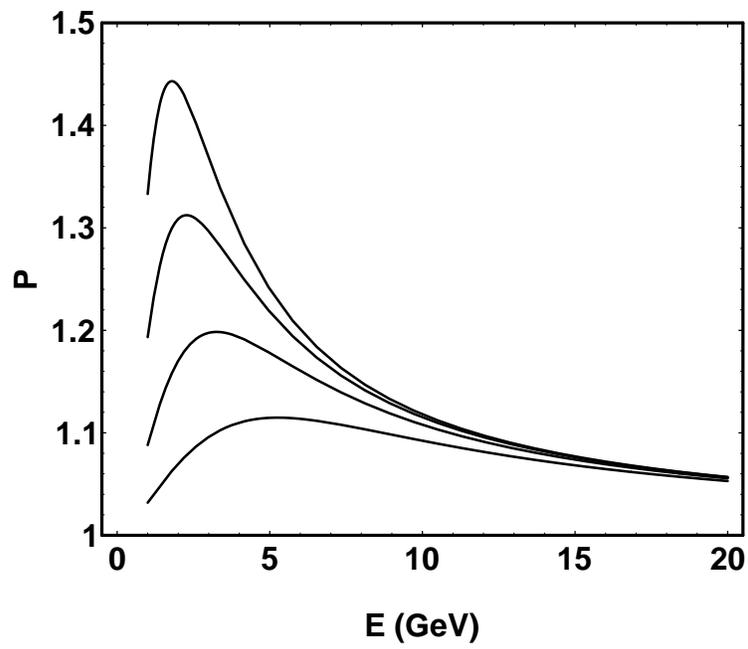
\end{document}